\documentstyle[twoside,jltp]{article}

\title{Ferromagnetism in Hubbard Models:\\ Low Density Route}

\author{E. M\"uller--Hartmann\address{Institut f\"ur Theoretische Physik,
Universit\"at zu K\"oln, Z\"ulpicher Str. 77, 50937 K\"oln, Germany}}

\runninghead{\bf E.~M\"uller--Hartmann}{Ferromagnetism in Hubbard Models}

\begin{document}

\begin{abstract}

Thirty years ago the Hubbard model was introduced by Gutzwiller, Hubbard and
Kanamori with the main purpose of mimicking the ferromagnetism of transition
metals. Soon after, Nagaoka and Thouless pointed out a basic mechanism for
ferromagnetism in strongly correlated electron systems by studying the motion
of a single hole in a half--filled Hubbard model. This important work was hoped
to shed light onto metallic ferromagnetism from the low doping regime.
Unfortunately, this low doping route towards ferromagnetism has not been
successful as far as rigorous results for finite doping concentrations are
concerned. In the work presented here, we start from the opposite limit of low
particle concentrations. In this limit we provide the first proof of a fully
polarized metallic ground state for a Hubbard model. The proof proceeds by
mapping Hubbard ``zigzag'' chains onto a continuum model with an
additional degree of freedom and local first Hund's rule coupling. For this
model the maximum total spin multiplet is shown to be the unique ground state
for infinite Hubbard coupling. Our proof may open a low density route towards
the understanding of the ferromagnetism of Hubbard models.

PACS numbers: 71.27.+a, 75.10.Lp
\end{abstract}

\maketitle

\vspace{0.3in}

The Hubbard model was introduced about 30 years ago \cite{gutz,hubb,kana}
mainly for the purpose of mimicking the ferromagnetism of transition metals
(see also the the paper by R. Strack and D. Vollhardt in this volume).
In view of this it is quite astonishing that up to this point our understanding
of the ferromagnetism of Hubbard models is very limited. All the rigorous
results proving the presence of a ground state with net polarization
\cite{lieb,mieltas,stravoll} are either concerned with
half--filled bands (i.e.~with ferromagnetism of insulators) or with somewhat
artificial flat--band systems which have ferromagnetic ground states even for
vanishing interaction strength. Very recently Tasaki succeeded in proving
local stability of the ferromagnetic ground state for nearly--flat--band
systems \cite{newtas}. There is not a single rigorous result which
would provide a positive answer towards the basic question which was in the
minds of the fathers \cite{gutz,hubb,kana} of the Hubbard model:
\begin{itemize}
\item Is a strong intra--atomic Coulomb repulsion sufficient for pushing a
partially filled (i.e.~paramagnetic) band into a fully polarized ferromagnetic
ground state (Nagaoka state)?
\end{itemize}

Nagaoka's theorem \cite{nag,thou} was, undoubtedly, an early important
contribution to the subject. It tells us the precise conditions for the
formation of ferromagnetic polarons around electrons or holes doped into a
half--filled Hubbard model at infinite coupling strength
(see also \cite{matt,tas,stravoll}). The puzzling results obtained for two and
more electrons or holes \cite{puzz} made obvious, though, that the problem of
the global alignment of ferromagnetic polarons, in case they existed, is
much harder to solve. In fact, not a single rigorous result establishing a
ferromagnetic ground state for finite doping concentration is known. The low
doping route towards metallic ferromagnetism which was so impressively opened
by Nagaoka's theorem has failed as far as rigorous results are concerned.

There are, however, several pieces of evidence in favor or in disfavor of
ferromagnetism for Hubbard models on various lattices. Let us first look at
the hypercubic lattices of dimensions $d=1$ to $d=\infty$. For the linear
chain ($d=1$) it is exactly known that there is never a ferromagnetic ground
state, except for the limit of infinite coupling strength where due to complete
spin degeneracy the fully polarized states are among the many ground states
\cite{liebwu}. In the limit of infinite dimension a variational argument
using Gutzwiller wave functions also proves that the ground state is never
fully polarized, not even for infinite coupling \cite{faz}. Here, partially
polarized ground states have not been excluded and probably exist, but have
not been established so far because the numerical solution of the mean--field
theory at $d=\infty$ (see the paper by A. Georges in this volume) is hard in
the very strong coupling limit. Variational estimates of the regime of
stability of a Nagaoka ground state suggest that the lower the dimension of the
lattice the larger this regime is, if it exists. Therefore, this
problem has been studied most extensively for the square lattice. The
variational proofs of instability of the Nagaoka state for the square lattice
leave a rather small regime of possible stability which can be characterized
by a critical electron or hole concentration (relative to half--filling) of
$\delta_{cr}=0.29$ and by a critical coupling of $U_{cr}=63t$ where $t$
denotes the nearest neighbor hopping amplitude \cite{ska,be,vdle,hmh}.
The estimates of the ground state energy deduced from a high temperature
expansion \cite{plo} allow for a critical density $\delta_{cr}$ (again relative
to half--filling)
of at most 0.10 and are consistent with $\delta_{cr}=0$. This result appears
not in agreement with finite square plaquette Lanczos data \cite{hir}
and with density matrix renormalization group (DMRG) extrapolations from
multi--chain systems \cite{lp} which both suggest a $\delta_{cr}$ of about
0.20. The evidence in the case of the square lattice seems still controversial
and not conclusive. The even more favorable case of the double chain has not
been studied as extensively as the square lattice. Lanczos data \cite{hir}
and DMRG data \cite{lp} agree on $\delta_{cr}=0.22$ for the double chain.
It would be very helpful if a high temperature expansion for the double chain
also became available.

In the presence of so much uncertainty in the case of hypercubic lattices it
should not be overlooked that there is positive evidence for Hubbard models
being much more fond of ferromagnetism on non--bipartite lattices
\cite{ska,mhhh}.
To avoid confusion in view of the absence of particle--hole symmetry for
non--bipartite lattices we assume the nearest neighbor hopping amplitude $t$
to be positive (which implies the band minimum to be found at the center of the
Brillouin zone). It is then demonstrated with rather simple variational wave
functions that for less than half--filling ($n<1$) the Nagaoka state is
never a ground state of the Hubbard model. For more than half--filling ($n>1$)
both variational estimates \cite{ska,mhhh} and Lanczos data \cite{hir} suggest
sizeable regimes with a Nagaoka ground state. This is perfectly consistent
with the Nagaoka theorem which predicts ferromagnetic polarons for $n>1$ only
\cite{nag}. The triangular lattice which is a prototype two--dimensional
non--bipartite lattice has been investigated with particular care. The best
variational calculations \cite{hkrmh} estimate the critical density at
infinite coupling $U$ as $n_{cr}\le1.887$, while Lanczos data appear to
confirm \cite{hir} that there is a Nagaoka ground state for $1<n<1.76$. For the
three--dimensional {\it fcc} and {\it hcp} lattices similar investigations
\cite{ska,mhhh,hir} suggest a Nagaoka ground state for the whole regime
$1<n<2$ at sufficiently large coupling. The same type of behavior follows from
Lanczos data for a quasi--one--dimensional lattice, the chain of triangles
\cite{hir}. We thus have good evidence for the rather striking phenomenon of
fully polarized ground states in the limit of small hole densities (relative
to complete filling, i.e.~to $n=2$). It is this phenomenon which we are going
to use in this paper as a starting point for opening a new route towards the
understanding of the ferromagnetism of Hubbard models. Since one can always map
the limit of small hole densities ($n\to2$) onto the limit of small particle
densities ($n\to0$), employing a particle--hole transformation, we will call
this route the ``low density route'' in contrast to what we called the ``low
doping route'' in the discussion above.

In what follows we will consider the $t-t'$ Hubbard chain
\begin{eqnarray}\label{H}
{\cal{H}} & = &
-\sum_{i\sigma} \left(t c_{i\sigma}^{\dagger}
c_{i+1\sigma}^{\vphantom{\dagger}}+
t' c_{i\sigma}^{\dagger} c_{i+2\sigma}^{\vphantom{\dagger}}+h.c.\right)
+U\sum_{i}n_{i\uparrow}n_{i\downarrow}.
\end{eqnarray}
For the special choice $t'=t$ this model describes the chain of triangles
mentioned above. We do prefer, in fact, to visualize this system as a zigzag
chain of atoms rather than as a linear chain because we will have to satisfy
the condition $t'>t/4$ to obtain a ferromagnetic ground state. This condition
can probably only be met if the second neighbor distance is shorter than twice
the first neighbor distance. With this in mind we also avoid any conceptual
conflict with the ``no--go--theorem'' for ferromagnetism in strictly
one--dimensional systems \cite{liebmat}.

As we want to focus on the low density limit of the zigzag chain we imagine
that a particle--hole transformation and a gauge transformation have been
performed such that from now on we will be interested in the case $t>0$ and
$t'<0$. Looking at the dispersion
\begin{eqnarray}\label{eps}
\epsilon(k) = -2t\cos k -2t'\cos 2k
\end{eqnarray}
of the $t-t'$ band we observe a qualitative change at the bottom of the band
as $t'$ decreases: the band minimum at $k=0$ splits into two equivalent
minima for $t'<-t/4$. We will see that in the low density limit the existence
of these two band minima is responsible for the formation of a Nagaoka ground
state in the strong coupling limit.

The two minima form around momenta $\pm k_0$ where
$\cos k_0 = -t/{4t'}$. At low densities only states in the neighborhood of
these two minima matter. It is important for the following argument that due to
the ultra--violet stability of one--dimensional systems which implies
intrinsic high energy cut--offs this is also true for the interacting system.
In the presence of two band minima the system therefore develops an
internal degree of freedom (left valley - right valley) at low densities.
Introducing the notation
$\tilde{c}^{\dagger}_{k\tau\sigma} := c^{\dagger}_{k-\tau k_0,\sigma}$
($\tau = \pm1$) and fitting cosine bands
$\tilde{\epsilon}_{\tau}(k) = \tilde{\epsilon}-2\tilde{t}\cos(k-\tau k_0)$
with $\tilde{\epsilon} = 2t'+t^2/{4t'}$ and $\tilde{t} = -4t'[1-(t/{4t'})^2]$
into the two minima of the original dispersion (\ref{eps})
we express the original Hamiltonian $\cal{H}$ in terms of new fields
distinguished by the symbol $\tilde{\vphantom{c}}\,$. The result of this simple
change of notation will be discussed using the new Hamiltonian
\begin{eqnarray}
\tilde{\cal{H}} & = & \tilde{\epsilon}N
-\tilde{t} \sum_{i\tau\sigma} \left(\tilde{c}_{i\tau\sigma}^{\dagger}
\tilde{c}_{i+1\tau\sigma}^{\vphantom{\dagger}}+h.c.\right)
+U\sum_{i\tau}\tilde{n}_{i\tau\uparrow}\tilde{n}_{i\tau\downarrow}\nonumber\\
& \  & -\tilde{J}\sum_{i}\left(\tilde{\bf S}_{i+}\cdot\tilde{\bf S}_{i-}-
\frac{1}{4}\tilde{n}_{i+}\tilde{n}_{i-}\right).
\end{eqnarray}
where $\tilde{\bf S}_{i\tau}$ and $\tilde{n}_{i\tau}$ denote the spin and
particle density operators of state $\tau=\pm$ at site $i$.

The Hamiltonian $\tilde{\cal{H}}$ describes two Hubbard chains with nearest
neighbor hopping $\tilde{t}$ which are coupled to each other by the on--site
exchange interaction $\tilde{J}$. It has twice as many degrees of freedom as
the original Hamiltonian (\ref{H}). Nevertheless, $\cal{H}$ and
$\tilde{\cal{H}}$ are
closely related in the limit of small particle densities: for $\tilde{J} = 2U$
the two models have one and the same continuum limit. The ferromagnetic
exchange interaction of strength $2U$ results from the inter--valley
scattering contained in the Hubbard term of the original model.

The presence of a ferromagnetic exchange interaction $\tilde{J}>0$ in
$\tilde{\cal{H}}$ suggests to expect ferromagnetic ground states for this
model. It is, in fact, possible to arrive at the following rigorous statement:
\begin{itemize}
\item At $U=\infty$, for any $\tilde{J}>0$ and for any particle number
$0<N<2L$ all the ground states of $\tilde{\cal{H}}$ have maximum total spin
$S = N/2$.
\end{itemize}

Here $L$ is the number of sites per chain such that $N = 2L$ corresponds to
half--filling of the double chain system. The proof of the above statement is
quite simple. Let us first consider the case $\tilde{J}=0$ where we have two
uncoupled Hubbard chains. From the exact solution \cite{liebwu} we know that
at $U=\infty$ the ground states are given by a Slater determinant of spinless
particles for each chain combined with a completely arbitrary spin wave
function. Since the kinetic energy is a convex function of the particle number
the particles distribute equally among the two chains ($N_+=N_-$ for even $N$
and $N_+=N_-\pm1$ for odd $N$).
Including now the exchange term we see that the energy of the
ground states is increased unless particles are in a triplet state whenever
they meet at the same site. The only way to make sure that particles never
meet in a singlet state is to have maximum total spin. This simple argument is
easily made rigorous by a formal proof that out of the subspace of completely
spin degenerate ground states of the uncoupled chains only those
with a totally symmetric spin wave function are uneffected by the positive
semi--definite exchange term.

The proof that $\tilde{\cal{H}}$ has no other than fully polarized ground
states is an interesting result in its own right. Interpreting the internal
degree of freedom as an orbital degeneracy we can rephrase this result by
saying that a Hubbard chain with orbital degeneracy and first Hund's rule
coupling has Nagaoka ground states for $U=\infty$. We are reminded of the fact
that orbital degeneracy combined with ferromagnetic Hund's rule coupling is a
genuine feature of real ferromagnets in the first place. Probably, omitting
orbital degeneracy in modeling metallic ferromagnets by the single band
Hubbard model \cite{gutz,hubb,kana} was an oversimplification after all. In
cases where the Hubbard interaction alone is not sufficient to drive a system
ferromagnetic Hund's rule coupling might serve to provide the final kick.

Despite of this remark we now turn back to the original single band model. Due
to the equivalence of models discussed above we conclude that the continuum
limit of $\cal{H}$ has only fully polarized ground states at $t'<-t/4$ and
$U=\infty$. This result represents the first proof of metallic ferromagnetism
in a Hubbard model. Applying a scale transformation to the continuum model we
find the following formula for the ground state energy per site at given
densities of particles $n$ and magnetization $m$:
\begin{eqnarray}
E_0/L = (\tilde{\epsilon}-2\tilde{t})n+n^3\tilde{t}\Phi(U/{n\tilde{t}},m/n).
\end{eqnarray}
We don't know the scaling function $\Phi(u,\mu)$ since the model is not
integrable. But from the above proof we infer
\begin{eqnarray}
\Phi(\infty,\mu)>\Phi(\infty,1)\quad(\mu<1).
\end{eqnarray}

The above result opens a low density route towards the understanding of the
ferromagnetism of Hubbard models on non--bipartite lattices. We have learned
that in the presence of several valleys at the bottom (or top, after a
particle--hole transformation) of the band the Hubbard interaction implies a
ferromagnetic exchange interaction between particles in the various valleys.
In one dimension this turned out to be sufficient for making the strong
coupling system ferromagnetic. It is presently not obvious how this result
can be generalized. Two valleys are also found in the band structure of the
triangular lattice. The discussion presented above tells us \cite{hkrmh} that
in this case two valleys are not sufficient for ferromagnetism at low
densities. Nevertheless, multi--valley situations seem to be a generic
prerequisite for Hubbard models having ferromagnetic ground states in a broad
range of particle densities. One may wonder whether the low density route
towards ferromagnetism will turn out to be more successful than the low doping
route.

\section*{ACKNOWLEDGMENTS}
This research is supported by SFB 341 of Deutsche Forschungsgemeinschaft.

\end{document}